\documentclass[conference]{IEEEtran}
\IEEEoverridecommandlockouts

\usepackage[english]{babel}
\usepackage[T1]{fontenc}
\usepackage{amssymb, amsmath, amsthm}
\usepackage{mathtools}
\usepackage{cite}
\usepackage{graphicx}
\usepackage[caption=false,font=footnotesize,farskip=0pt]{subfig}
\usepackage{xcolor}
\usepackage{multirow}
\usepackage[ruled,vlined,linesnumbered]{algorithm2e}
\usepackage{enumitem}
\usepackage{stfloats}
\usepackage[normalem]{ulem}
\usepackage{stix}
\usepackage{booktabs}
\usepackage{tabularx}
\usepackage{multirow}
\usepackage{multicol}

\usepackage{xcolor, colortbl}
\usepackage[acronym,,nohypertypes={acronym,notation}]{glossaries}
\newacronym{3d}{3D}{three dimensional}

\newacronym[plural=ACKs]{ack}{ACK}{acknowledgment}
\newacronym{aoa}{AoA}{angle of arrival}
\newacronym{awgn}{AWGN}{additive white Gaussian noise}
\newacronym{aod}{AoD}{angle of departure}
\newacronym{au}{AU}{active user}

\newacronym{rat}{RAT}{reflected-angular training}
\newacronym[plural=APs, firstplural=access points (APs)]{ap}{AP}{access point}
\newacronym{b5g}{B5G}{Beyond-5G}
\newacronym[plural=BSs, firstplural=base stations (BSs)]{bs}{BS}{base station}

\newacronym{cc}{CC}{control channel}
\newacronym{ce}{CE}{configuration estimation}
\newacronym{chest}{CHEST}{channel estimation}
\newacronym{csi}{CSI}{channel state information}
\newacronym{cdf}{cdf}{cumulative distribution function}
\newacronym{crc}{CRC}{cyclic redundancy check}
\newacronym{crlb}{CRLB}{Cram\'er-Rao lower bound}
\newacronym{ccdf}{CCDF}{complementary cumulative distribution function}

\newacronym{dc}{DC}{direct current}
\newacronym{dsp}{DSP}{digital signal processing}
\newacronym{dl}{DL}{downlink}
\newacronym{doa}{DoA}{direction-of-arrival}
\newacronym{emf}{EMF}{electromagnetic field}
\newacronym{em}{EM}{electromagnetic}
\newacronym{fp}{FP}{fractional program}
\newacronym{glrt}{GLRT}{generalized likelihood ratio test}
\newacronym[plural=HRISs, firstplural=Hybrid Reconfigurable Intelligent Surfaces (HRISs)]{hris}{HRIS}{hybrid reconfigurable intelligent surface}
\newacronym{iid}{i.i.d.}{independent and identically distributed}
\newacronym{ios}{IoS}{Internet-of-Surfaces}
\newacronym{iot}{IoT}{Internet-of-Things}

\newacronym[first=OnRMap]{our}{OnRMap}{}

\newacronym[plural=KPIs, firstplural=key performance indicators (KPIs)]{kpi}{KPI}{key performance indicator}
\newacronym{ls}{LS}{least-squares}
\newacronym{lf}{LF}{low frequency}
\newacronym[plural=LISs, firstplural=large intelligent surfaces (LISs)]{lis}{LIS}{large intelligent surface}
\newacronym{los}{LoS}{line-of-sight}

\newacronym{mac}{MAC}{medium access control}
\newacronym{mf}{MF}{matched filter}
\newacronym{mimo}{MIMO}{multiple-input multiple-output}
\newacronym{mmimo}{M-MIMO}{massive MIMO}
\newacronym{miso}{MISO}{multiple-input single-output}
\newacronym{ml}{ML}{machine learning}
\newacronym{mle}{ML}{maximum-likelihood estimator}
\newacronym{mmse}{MMSE}{minimum mean squared error}
\newacronym{mmtc}{mMTC}{massive machine-type communications}
\newacronym{mrc}{MRC}{maximum-ratio combining}
\newacronym{mse}{MSE}{mean-squared error}

\newacronym{nlos}{NLoS}{non-line-of-sight}

\newacronym{pe}{PE}{passive element}
\newacronym[plural=PDFs]{pdf}{PDF}{probability distribution function}
\newacronym{pla}{PLA}{planar linear array}
\newacronym{pap}{P\&P}{plug-and-play}
\newacronym{ppp}{PPP}{Poisson point process}

\newacronym{ra}{RA}{random access}
\newacronym[plural=RMs]{rm}{RM}{radio map}
\newacronym{rmap}{RMap}{radio mapping}
\newacronym{rap}{RAP}{random access procedure}
\newacronym[plural=RISs, firstplural=reconfigurable intelligent surfaces (RISs), first=RIS]{ris}{RIS}{reconfigurable intelligent surface}
\newacronym{rf}{RF}{radio frequency}
\newacronym{rmse}{RMSE}{root-mean-square error}
\newacronym{rss}{RSS}{received signal strength}
\newacronym{rpca}{RPCA}{robust principal component analysis}
\newacronym{bw}{B\&W}{black-and-white}
\newacronym{2d}{2-D}{two dimensional}
\newacronym{dbscan}{DBSCAN}{Density-based spatial clustering of applications with noise}
\newacronym[plural=MCSs]{mcs}{MCS}{Monte Carlo simulation}

\newacronym{se}{SE}{squared error}

\newacronym{sdp}{SDP}{semidefinite programming}
\newacronym{sdr}{SDR}{semidefinite relaxation}
\newacronym{sic}{SIC}{successive interference cancellation}
\newacronym{sinr}{SINR}{signal-to-interference-plus-noise ratio}
\newacronym{smse}{SMSE}{sum mean squared error}
\newacronym{sdma}{SDMA}{space-division multiple-access}
\newacronym{snr}{SNR}{signal-to-noise ratio}
\newacronym{soa}{SoA}{state-of-the-art}
\newacronym{sre}{SRE}{smart radio environment}

\newacronym{toa}{ToA}{time-of-arrival}

\newacronym{tdm}{TDM}{time-division multiplexing}
\newacronym{tdma}{TDMA}{time-division multiple access}
\newacronym{tdd}{TDD}{time-division duplex}
\newacronym{tem}{TEM}{transverse electromagnetic mode}

\newacronym{uatf}{UatF}{use-and-then-forget}
\newacronym[plural=UEs, firstplural=users' equipment (UEs)]{ue}{UE}{user's equipment}
\newacronym{ul}{UL}{uplink}
\newacronym{ula}{ULA}{uniform linear array}
\newacronym{upa}{UPA}{uniform planar array}

\newacronym{mr}{MR}{Maximal-ratio}
\newacronym{5G}{5G}{5th Generation}
\newacronym{rti}{RTI}{radio tomographic imaging}
\newacronym{wsn}{WSN}{wireless sensor network}

\definecolor{cian}{rgb}{.02,.7,.95}
\definecolor{gold}{rgb}{0.85,.66,0}

\newcommand{\colr}{\textcolor{red}}
\newcommand{\colnb}{\textcolor{green!50!black!80}}

\newcommand{\bUp}{\boldsymbol{\Upsilon}}

\newcommand{\colv}{\textcolor{violet}}

\DeclareMathOperator{\vect}{vec}

\title{{OnRMap: An Online Radio Mapping Approach for Large Intelligent Surfaces}
\thanks{
This work was supported in part by the Coordenação de Aperfeiçoamento de Pessoal de Nível Superior - Brasil (CAPES) – Finance Code  001 and by the National Council for Scientific and Technological Development (CNPq) of Brazil under Grants 405301/2021-9, 141485/2020-5, and 310681/2019-7. Victor Croisfelt, and Petar Popovski were supported by the Villum Investigator Grant “WATER” from the Velux Foundation, Denmark, and partly by the H2020 RISE-6G project financed by the European Commission under grant no. 101017011. Cristian J. Vaca-Rubio was supported by the European Union's Horizon EUROPE research and innovation program under grant agreement No. 101037090 - project CENTRIC.
}
}

\author{
    \IEEEauthorblockN{
        {Herman L. dos Santos}\IEEEauthorrefmark{1},
        {Victor Croisfelt}\IEEEauthorrefmark{2},
        {Cristian J. Vaca-Rubio}\IEEEauthorrefmark{2},
        {Taufik Abrão}\IEEEauthorrefmark{1},
        and {Petar Popovski}\IEEEauthorrefmark{2} \\
        }
        \IEEEauthorblockA{
        \IEEEauthorrefmark{1}\textit{Department of Electrical Engineering, Universidade Estadual de Londrina}, Londrina, Brazil\\
        \IEEEauthorrefmark{2}\textit{Department of Electronic Systems, Aalborg University}, Aalborg, Denmark\\
        E-mail: hermanlds@gmail.com, \{vcr, cjvr, petarp\}@es.aau.dk, and taufik@uel.br
        }
}

\begin{document}

\maketitle

\begin{abstract}
We introduce OnRMap, an online radio mapping (RMap) approach for the sensing and localization of active users (AUs), devices that are transmitting radio signals, and passive elements (PEs), elements that are in the environment and are illuminated by the AUs' radio signals. OnRMap processes the signals received by a large intelligent surface and produces a radio map (RM) of the environment based on signal processing techniques. The method then senses and locate the different elements without the need for offline scanning phases, which is important for environments with frequently changing spatial layouts. Empirical results demonstrate that OnRMap presents a higher localization accuracy than an offline method, but the price paid for being an online method is a moderate reduction in the detection rate.
\end{abstract}


\section{Introduction}
\Gls{lis} is an important concept on the evolution path of wireless multi-antenna systems. It originally refers to a continuous electromagnetic surface able to transmit and receive radio waves\cite{Hu2018}. In practice, a \gls{lis} is envisioned as a collection of closely-spaced antenna elements deployed across a large 2D surface, which can be easily integrated into the propagation environment, \emph{e.g.}, placed on walls or ceilings. In addition to its well-investigated communication capabilities \cite{Dardari2019, Hu2018}, \gls{lis} also holds potential for radio sensing and localization~\cite{vaca2021assessing}, with applications in self-driving vehicles, unmanned aerial vehicles, or autonomous robots. These use cases normally require the construction of \glspl{rm}, whose process can exploit the \gls{rf} signals emitted by the wireless devices in the environment of interest, termed \glspl{au}. Moreover, static and/or dynamics objects that are not transmitting \gls{rf} signals, termed \glspl{pe}, can also be detected/sensed and located by exploiting the multipath components of the \gls{rf} signals transmitted by the \glspl{au}. We refer to \gls{rmap} as the process of obtaining \glspl{rm}, which is the main subject of this paper. 

In \cite{Vaca-Rubio2021,vaca2022floor}, the authors used a \gls{lis} to obtain \gls{rm}, treating the received signals at the \gls{lis} as a digital image and creating an \gls{rmap} method based on techniques from digital image processing and computer vision. Despite the good detection performance of \glspl{au} and \glspl{pe}, digital image processing requires offline processing, which is not suitable for environments with frequently changing spatial layouts. Other previous \gls{rmap} methods rely on the discretization of outdoor \cite{Tong2022} and indoor \cite{Tong2021} static environments, evaluating the path between \glspl{au} and \glspl{pe} in a pixel-like manner. The 
critical problem with pixel-based approaches is that good detection performance requires increased pixel granularity, resulting in in exponential  complexity. Differently from this, the authors of \cite{wilson2010radio, zhao2018through} used the related concept of \gls{rti} that allowed them to obtain an \gls{rm} of moving \glspl{pe} (humans) by imaging their attenuation in a \gls{wsn} comprised of fixed-located sensors in a square area. However, this  demands dedicated sensors, making it more expensive than methods that exploit widespread wireless \glspl{au}.

This work proposes \textit{\gls{our}}, an online \gls{rmap} approach based on classical signal processing techniques. In contrast to \cite{Vaca-Rubio2021, vaca2022floor}, \gls{our} eliminates the need for offline scanning phases, being more robust to dynamic environments at the cost of a reasonably lower detection rate in comparison to \cite{Vaca-Rubio2021}. The numerical results indicate that \gls{our} provides a higher localization accuracy of the \glspl{pe}.

\section{System Model}
\begin{figure}[t]
    \centering
    \includegraphics[trim=0mm 12mm 12mm 0mm, clip, width=.7\linewidth]{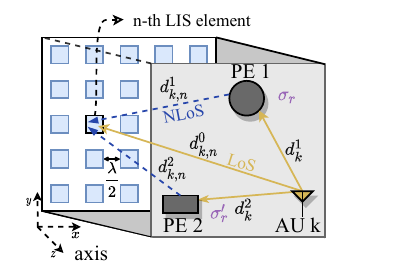}
    \caption{Indoor communication scenario with a \gls{lis} installed on the ceiling. The \gls{los} (\gls{au}-to-\gls{lis}) and \gls{nlos} (\gls{au}-to-\glspl{pe}-to-\gls{lis}) rays from a single \gls{au} are depicted. \glspl{pe} can be different objects characterized by different parameters $\sigma_r$.}
    \label{fig:floorplan}
\end{figure}

Consider the indoor communication system within an enclosed room and where a \gls{lis} is placed on the ceiling, as illustrated in Fig. \ref{fig:floorplan}. The \gls{lis} covers the whole room beneath it and is composed of a \gls{upa} containing $N=N_x \cdot N_y$ antenna elements equally spaced by $\lambda/2$. Within the room, there are $R$ \glspl{pe} and $K$ \glspl{au}. 

\subsection{Channel Model}
Assume that when the $K$ \glspl{au} transmit \gls{ul} signals, those impinge at the \gls{lis} either as a result of \gls{los} and/or as \gls{nlos} propagation. The latter occurs due to reflections of the transmitted signals on the \glspl{pe} present in the room. For mathematical tractability, we ignore the \gls{nlos} components resulting from reflections on the wall and floor in the formulation; besides, we assume one reflection per \gls{pe}. Let $\mathbf{h}_k \in \mathbb{C}^{N \times 1}$ denote the channel vector of the $k$-th \gls{au} to the \gls{lis}. Considering an indoor scenario, we use the spherical wavefront assumption 
and adopt the Saleh-Valenzuela model \cite{SalehValenzuela1987}, where the $n$-th entry of $\mathbf{h}_k$ is
\begin{equation}
    h_{k,n} = \underbrace{\vphantom{\sum\limits_{r=2}^{R}} \beta_{k,n}^{0} e^{-j\frac{2\pi}{\lambda} d_{k,n}}}_{\text{LoS}} + \underbrace{\sum\limits_{r=1}^{R}\beta_{k,n}^{r} e^{-j\frac{2\pi}{\lambda} d^{r}_{k,n}}}_{\text{NLoS}},
    \label{eq:channelModelh}
\end{equation}
with $\lambda$ being the carrier wavelength and the {superscript $[\cdot]^r$} denoting the $r$-th multipath component from a total of $R+1$ components; specially, $\beta^{0}_{k,n} \in \mathbb{R}^+$ denotes the \gls{los} component. The Euclidean distance from the \gls{au} to the $n$-th \gls{lis} antenna is denoted by $d^{0}_{k,n}$, while $d^{r}_{k,n}$ denotes the distance of the $n$-th \gls{lis}' element to the $r$-th \gls{pe}. The channel gain of the $r$-th multipath component can be modeled as \cite{goldsmith2005}:
\begin{equation}
    \beta^r_{k,n} = \frac{\lambda}{4 \pi}\frac{\sqrt{\sigma_r} e^{-j\Delta \phi_{r,n}}}{(d^r_{k} + d^r_{k,n})}
    \label{eq:complexPathlossBeta}
\end{equation}
with $d^r_{k}$ representing the distance between the $k$-th \gls{au} and the $r$-th \gls{pe}, $\sigma_r$ denoting the reflection loss by the $r$-th \gls{pe}, and $\Delta \phi_{r,n} = 2 \pi({d^r_{k} + d^r_{k,n} - d^0_{k,n}})/{\lambda}$ is the phase difference (Doppler shift) between the \gls{los} and the $r$-th \gls{nlos} components. The parameter $\sigma_r$ models a random variable subject to conductivity, relative permittivity, and permeability of the {\glspl{pe}}, whose value can differ for different types of \glspl{pe}.
%

\section{Radio Mapping: An Overview}\label{sec:rmOverview}

Suppose the \gls{chest} phase of a \gls{lis} system. 
The $K$ \glspl{au} transmit using $K$ orthogonal pilots such that the received signal $\mathbf{Y}\in\mathbb{C}^{N\times K}$ at the \gls{lis} is:
\vspace{-2mm}
\begin{equation}
   \mathbf{Y} =  p \mathbf{H} + \mathbf{W}, 
   \label{eq:receivedSignaly}
\end{equation}
where $p$ is the \gls{ul} transmit power, which is assumed to be equal to all the \glspl{au}, $\mathbf{H}=[\mathbf{h}_1,\mathbf{h}_2,\dots,\mathbf{h}_K]\in\mathbb{C}^{N\times K}$ is the channel matrix, and $\mathbf{W}\in\mathbb{C}^{N\times K}$ is the receiver noise matrix whose entries are \gls{iid} according to $\mathcal{CN}({0},\sigma^2_w)$. To exploit $\mathbf{Y}$ for \gls{rmap}, a spatial \gls{mf} $\hat{\mathbf{h}}\in\mathbb{C}^{N_f\times 1}$ was proposed in \cite{Vaca-Rubio2021,vaca2022floor}, whose filter coefficients are computed as:

\begin{equation}
    \hat{h}_n = {t_n}e^{-j \frac{2\pi}{\lambda}{d_n}}, \quad \forall n\in\{1,\dots, N_f\},
    \label{eq:sphCoeff}
\end{equation}
where $t_n$ is the $n$-th element of a weighting vector $\mathbf{t}\in {\mathbb{R}_+}^{N_f\times 1}$, $e^{-j \frac{2\pi}{\lambda}{d_n}}$ defines a spherical wave phase-matching component, adjusted to the Euclidean distance $d_n = ({d_{x,n}^2 + d_{y,n}^2 + d_{z,n}^2})^{1/2}$ between a reference $(x,y,z)$ point in space and the $n$-th \gls{lis} antenna, and $N_f \leq N$ is the number of taps of the filter. Let denote the $k$-th column of $\mathbf{Y}$ in Eq. \eqref{eq:receivedSignaly} as $\mathbf{Y}_k\in\mathbb{C}^{N_x\times N_y}$ with $\mathbf{y}_k=\vect(\mathbf{Y}_k)$. We also let $\underline{\hat{\mathbf{h}}}=[\hat{\mathbf{h}},\mathbf{0}]\in\mathbb{C}^{N\times 1}$ be the \gls{mf} after zero-padding according to $\max{(N_f,N)}$. Then, the contribution ${\boldsymbol{\Upsilon}}_k\in\mathbb{C}^{N_x\times N_y}$ to the primary \gls{rm} from the $k$-th \gls{au} is obtained as follows:
\begin{equation}
    {\boldsymbol{\Upsilon}}_k = \underline{\hat{\mathbf{H}}} {{\otimes} }\mathbf{Y}_k, \label{eq:RMFiltering}
\end{equation}
where $\underline{\hat{\mathbf{h}}}=\vect(\underline{\hat{\mathbf{H}}})$. The primary \gls{rm} ${\boldsymbol{\Upsilon}}\in\mathbb{C}^{N_x\times N_y}$ is \cite{Vaca-Rubio2021}:
\begin{equation}\label{eq:sumRM}
    \bUp = {\left(\sum\limits_{k=1}^K \mathbf{Y}_k \right) {\otimes} \underline{\hat{\mathbf{H}}}}.
\end{equation}
When designing the filter, the authors of \cite{Vaca-Rubio2021} considered the components $d_{x,n}$ and $d_{y,n}$ of $d_n$ to be the distances of each $n$-th \gls{lis} element to the center of the room, while the focal height $d_{z,n}$ is a parameter subject to design. Further processing can be cast over $\bUp$ to perform sensing and localization.

We highlight two main drawbacks in the pre-processing from~\cite{Vaca-Rubio2021}. First, by using the combination in Eq. \eqref{eq:sumRM}, the orthogonality among the signals from different \glspl{au} is lost. This now non-orthogonal superposition may incur loss of relevant information, as the \glspl{pe} are being irradiated from different angles. This can produce different Dopplers' shifts, as indicated by Eq. \eqref{eq:complexPathlossBeta}, which can in turn mitigate the contribution of the \gls{nlos} components of interest. In other words, the contribution of the \glspl{pe} can be \textit{overshadowed} in the \glspl{rm}. Second, the filter structure is capable of matching the phase of the \gls{los} components, but outputs some \textit{distortion}, especially in the neighborhood where the \gls{los} ray impinges. 

\begin{figure}
    \centering
    \includegraphics[trim= 0mm 5mm 0mm 0mm, clip, width=.49\textwidth]{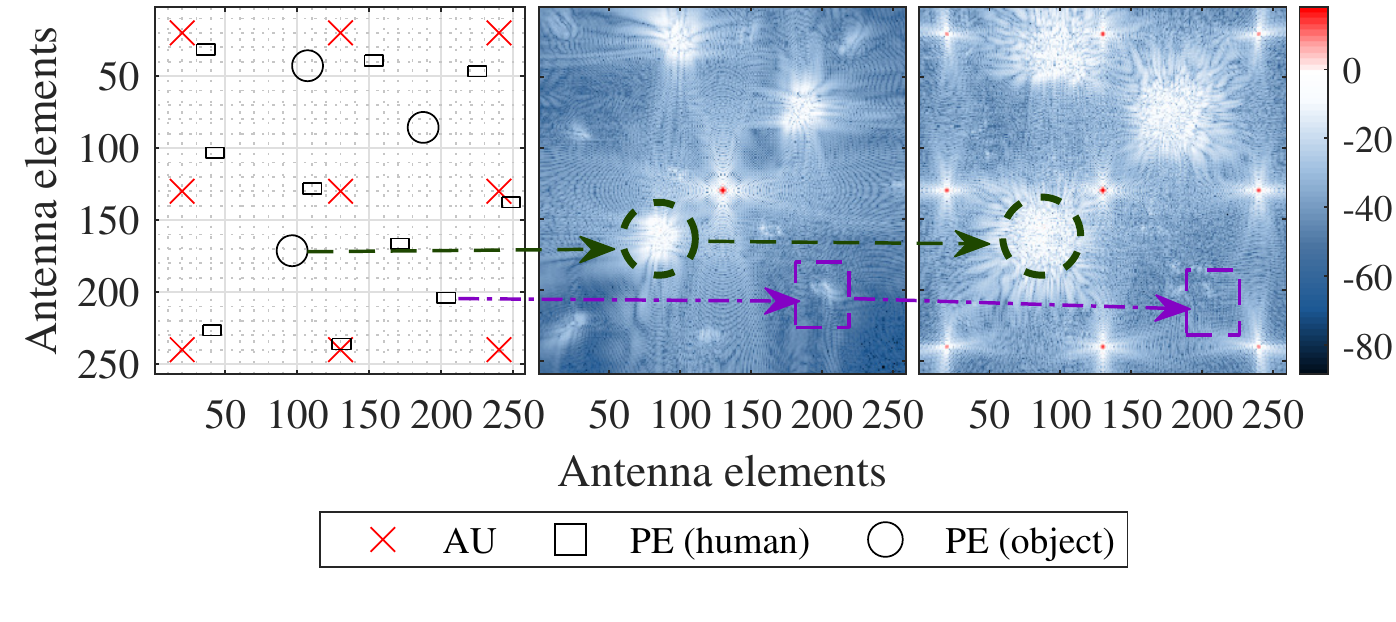}
    \caption{\glspl{rm} showing one drawback of the \gls{mf} used in \cite{Vaca-Rubio2021} in the indoor scenario specified in Appendix \ref{appx:illustrative-indoor} with $K=9$ \glspl{au} and $R=13$ \glspl{pe}, where $R_{\rm obj}=3$ are metallic objects and $R_{\rm hum}=10$ are humans. \textit{Left}: ground-truth. \textit{Center}: single-\gls{au} $K=1$; \textit{Right}: superposed $K=9$ \glspl{au}. When the signals are superposed, it is noticeable that \glspl{pe} with higher reflection loss (humans w/ \colv{violet} line) get occluded, while the others (objects w/ \colnb{green} line) got enhanced.}
    \label{fig:diffRM}
\end{figure}

To illustrate, we consider the scenario described in Appendix \ref{appx:illustrative-indoor}, where we have $K=9$ \glspl{au} and $R=13$ \glspl{pe} with $R_{\rm obj}=3$ of them being cylindrical metallic objects and $R_{\rm hum}=10$ humans. Fig. \ref{fig:diffRM} contains (a) the ground truth \gls{rm}, where the \glspl{au} are represented by red crosses, while the \glspl{pe} by the geometrical forms -- metallic objects by the circles and the humans by the rectangles -- (b) the \gls{rm} of a single \gls{au}; and (c) the \gls{rm} of superposed signals of the $K=9$ \glspl{au}. The \gls{mf} filter used was the same as in \cite{Vaca-Rubio2021}, with ${t}_n = 1/d_n$ computed as in \eqref{eq:sphCoeff}. Fig. \ref{fig:diffRM}(b) exposes the effect of the distortion around the \gls{los} components with power as high as the rays reflected by \glspl{pe}. Fig.~\ref{fig:diffRM}(c) shows the behavior when the signals are superposed. Specifically, we see how the metallic objects are highlighted while the humans get occluded, making it more difficult to correctly detect the latter.
%
\section{OnRMap}\label{sec:onrmap}
Here we present \gls{our}, an online \gls{rmap} approach based on classical signal processing theory, which removes the need for offline scanning phases from~\cite{Vaca-Rubio2021,vaca2022floor}. We detail the four stages that comprise \gls{our} until we sense and locate the \glspl{au} and \glspl{pe}, with an emphasis on the \glspl{pe}, which are the most challenging to be detected. 

\subsection{OnRMap: An Overview}
\gls{our} consists of {four} steps, illustrated in Fig. \ref{fig:diagram}. 

\noindent \textbf{Step 0.} \textit{Signal Acquisition and Primary \gls{rm}.} In this initial step, we obtain the contributions $\bUp_k$ from the \glspl{au} $\forall k \in \{1,2, \dots, K\}$ to the primary \gls{rm} by filtering the signals $\mathbf{Y}_k$ with the \gls{mf} given in Eq. \eqref{eq:RMFiltering}. {D}ifferent from \cite{Vaca-Rubio2021} and based on the discussion made in Section \ref{sec:rmOverview}, we empirically chose the weight vector $\mathbf{t}$ using a \gls{2d} Taylor window \cite{Brookner1991}. This selection was motivated by the empirical reduction of filtering distortion and numerical improvement in the ratio between the \gls{nlos} components. The output of this step is the matrix $ \bar{\bUp} \in \mathbb{R}^{N \times K}$ containing the $K$ primary \glspl{rm} stacked in columns, \textit{i.e.}, $\bar{\bUp}= [ |\vect(\bUp_1)|, |\vect(\bUp_2)|, \dots \ |\vect(\bUp_K)|]$, where $|\cdot|$ here denotes element-wise absolute value.

\noindent \textbf{Step 1.} \textit{Estimation of the \gls{los} and \gls{nlos} Components.} Through this step, the primary \gls{rm} in $\bar{\bUp}$ is the input to the \gls{rpca} algorithm \cite{Candes2011}, which outputs the matrices $\hat{\bar{\bUp}}^{\rm LoS}$ and $\hat{\bar{\bUp}}^{\rm NLoS}$ corresponding to the estimations of the \gls{los} and \gls{nlos} components of $\bar{\bUp}$, respectively. We employ this method based on the observation that the \gls{nlos} components have similarities among themselves, \textit{e.g.}, the range of the power gain and location
{. In contrast,} \gls{los} components are a few data points with high power gains that are normally far apart. Hence, we can interpret the \gls{nlos} components to be low-rank components of $\bar{\bUp}$, whereas the \gls{los} {elements} are sparse. Thus, \gls{rpca} becomes useful, since it is a low-complexity method for estimating the low-rank components of a matrix. Based on the special focus on sensing and locating the \glspl{pe}, this block outputs the low-rank estimation, $\tilde{\bUp}^{\rm NLoS} \in \mathbb{R}^{N_y \times N_x}$.

\noindent \textbf{Step 2.} \textit{Separation of the \gls{nlos} Components.} This step translates the \gls{nlos} estimation from the previous step into data points for the inference step. To do so, we input $\vect(\tilde{\bUp}^{\rm NLoS})$ in k-means clustering \cite{Arthur2007}, which separates the data in two clusters that represent \textit{foreground} ( high power) and \textit{background} (low power) classes differentiate by their power level. The foreground forms shapes in \gls{2d} space and their perimeters are estimated with the Moore-Neighbor boundaries estimation algorithm \cite{Gonzalez2004}. The points that enclose the perimeters are stored in subsets that compound the set $\mathcal{B}$ and the total power levels each shape comprises in $\tilde{\bUp}^{\rm NLoS}$ are stored in the set $\mathcal{E}$, constituting the output of this step.

\noindent \textbf{Step 3.} \textit{Sensing and Localization Inference.} In this final block, we use the data in $\mathcal{B}$ and $\mathcal{E}$ to infer whether each subset in $\mathcal{B}$ is a type of \gls{pe} or just noise. {If it is a \gls{pe}, we can further identify which type of object it is, \emph{e.g.}, a human or a metallic object, and their positions are also inferred.} To do so, we employ \gls{dbscan} clustering {method} \cite{Ester1996dbscan} to cluster the data in $\mathcal{B}$ based on the distance between samples. It is also assigned to each cluster its total power by looking at the set $\mathcal{E}$. {The classification of each cluster on which type of \gls{pe} {(metallic object or human)} follows a decision rule based on the power each cluster has, that is, {lower and upper bounds are defined, and the clusters that fit in between are considered humans, those above are considered as objects, while the rest is noise.}

\begin{figure}[!t]
    \centering
    \includegraphics[trim={14mm 0mm 5mm 1mm},clip, width=.49\textwidth]{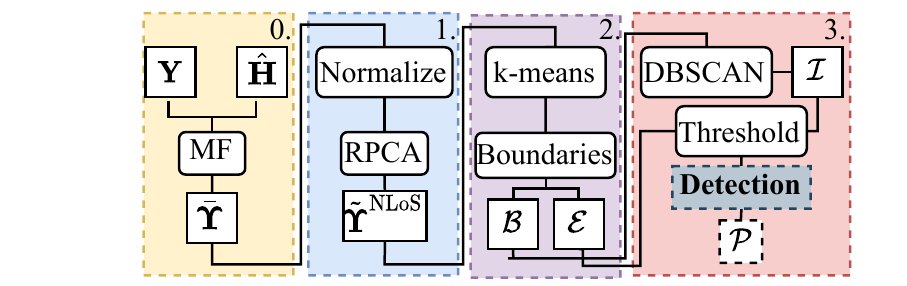}
    \caption{\gls{our} architecture.}
    \label{fig:diagram}
\end{figure}

\subsection{OnRMap: Detailed Description}
Below, we give further details on the other steps apart from Step 0, which was already detailed in \eqref{eq:RMFiltering}.




%
\subsubsection{Estimation of the LoS and NLoS Components}
\gls{rpca} solves the following optimization problem \cite{Candes2011}:
\begin{align}\label{eq:rpcaoptm}
    \min  \quad & \| \mathbf{L}\|_* + \lambda_{\rm RPCA} \| \mathbf{S} \|_1 \nonumber \\
    \mathrm{s.t.} \quad & \mathbf{L} + \mathbf{S} = \mathbf{M},
\end{align}
\noindent where $\mathbf{M}$ is the observation matrix, $\mathbf{L}$ and $\mathbf{S}$ are estimations of the low-rank and sparse components of $\mathbf{M}$, respectively, $\|\cdot \|_*$ and $\|\cdot\|_1$ are the nuclear and $\ell_1$ norm of a matrix, respectively, and $\lambda_{\rm RPCA}$ is a scalar. In our case, from Step 0 we have the matrix $\bar{\bUp} \in \mathbb{R}^{K\times N}$, which is obtained by stacking and normalizing the $\vect({\bUp}_k)$, $\forall k \in \{1, 2, \dots, K\}$. Then, we decompose it as:
\begin{equation}
    \bar{\bUp} = \bar{\bUp}^{\rm NLoS} + \bar{\bUp}^{\rm LoS}.    
\end{equation}
Define $\hat{\bar{\bUp}}^{\rm NLoS}$ and $\hat{\bar{\bUp}}^{\rm LoS}$ as estimations of $\bar{\bUp}^{\rm NLoS}$ and $\bar{\bUp}^{\rm LoS}$, respectively. Then, substituting in Eq. \eqref{eq:rpcaoptm}, we have
\begin{align}\label{eq:rpcanotation}
    \min  \quad & \lVert{\hat{\bar{\bUp}}}^{\rm NLoS}\rVert_* + \lambda_{\rm RPCA} \lVert\hat{\bar{\bUp}}^{\rm LoS}\rVert_1 \nonumber \\
    \mathrm{s.t.} \quad & \hat{\bar{\bUp}}^{\rm NLoS} + \hat{\bar{\bUp}}^{\rm LoS} = \bar{\bUp},
\end{align}
placing the separation of \gls{los} and \gls{nlos} components as equivalent to the optimization problem in \eqref{eq:rpcaoptm}. {The output of the algorithm $\hat{\bar{\bUp}}^{\rm NLoS}\in \mathbb{R}^{K\times N}$ is summed row-wise and reshaped to $\tilde{\bUp}^{\rm NLoS} \in \mathbb{R}^{N_y\times N_x}$; the same for the \gls{los}-related matrix.}

To illustrate how well \gls{rpca} performs, Fig. \ref{fig:rpca} shows the recovered low-rank {(left)} and sparse matrix {(right).} As can be seen, \gls{rpca} {can} estimate satisfactorily the \gls{nlos} {(left)} and \gls{los} {(right)} components. On the other hand, we can point out two disadvantages of this technique. First, although there is a reference value for $\lambda_{\rm RPCA}$, it may have to be tuned according to the scenario. Second, the problem of estimating low-rank and sparse components of matrices is NP-hard. \gls{rpca} solves a convex relaxed problem, incurring some information loss. However, most of the time, this method {estimates} satisfactory {the components} within less than twenty iterations. Despite that, for some realizations, part of the \glspl{pe} was outputted in the sparse (\gls{los}) component.

\begin{figure}[t!]
    \centering
    \includegraphics[trim = 0mm 0mm 0mm 0mm, clip, width=.82\linewidth]{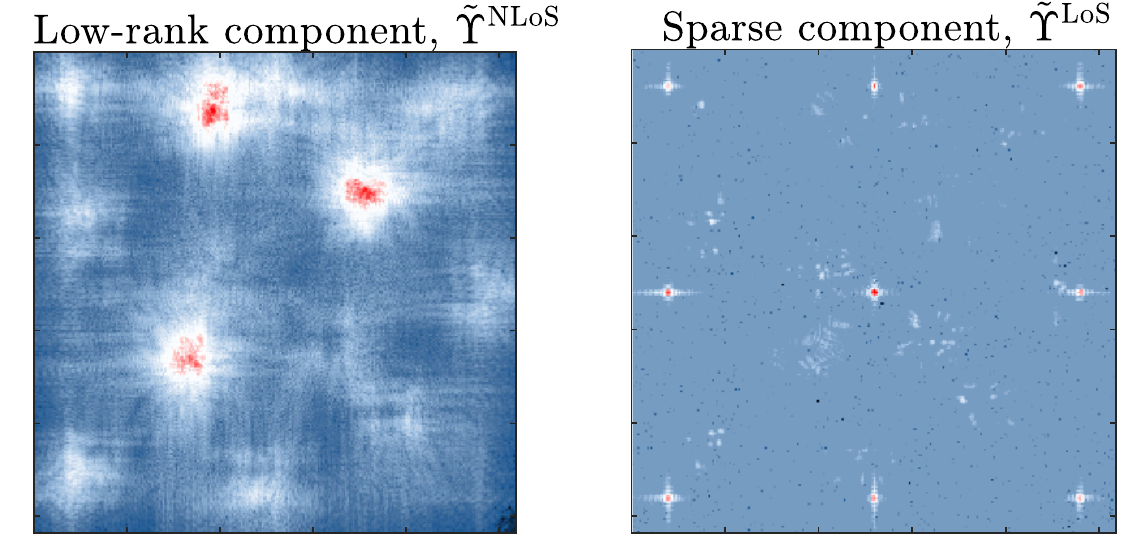}
    \caption{RPCA output. \textit{Left:} low-rank matrix $\hat{\bar{\bUp}}^{\rm NLoS}$ denoting the \gls{nlos} components. \textit{Right}: sparse matrix $\hat{\bar{\bUp}}^{\rm LoS}$ denoting the \gls{los} components. We set the parameter $\lambda_{\rm RPCA} = 3{N}^{-1/2}$ based on \cite{Candes2011} and experimentation.}
    \label{fig:rpca}
\end{figure}


%
\subsubsection{Separation of the NLoS Components}
First, let's define the binary k-means clustering operation as $\mathfrak{K}: \mathbb{R}^{N\times 1} \mapsto \{0,1\}^{N\times1}$. Then, we perform {entry-wise} $\vect (\bUp^{\rm km}) = \mathfrak{K}(\vect(\tilde{\bUp}^{\rm NLoS}))$, with $\bUp^{\rm km} \in \{0,1\}^{N_y \times N_x}$ being the matrix with the class of each data point in $\tilde{\bUp}^{\rm NLoS}$. In particular, we name class 0 as \textit{background} containing points with null to very low power and class 1 as \textit{foreground} containing components above a certain power threshold. We illustrate the output of the k-means $\bUp^{\rm km}$ on the left side of Fig. \ref{fig:segandbwbound}. Note the creation of certain regions that can be interpreted as geometrical shapes. Visually inspecting and comparing the shapes with the ground truth, {Fig. \ref{fig:diffRM}(a)}, we can infer that components that represent the metallic objects are more likely to be clustered in a larger and denser area, while the humans' ones are smaller and sparser.

Based on the observation made above, we employ the Moore-Neighbor boundaries estimation algorithm \cite{Gonzalez2004} to leverage these regions to detect the \glspl{pe}. The boundary estimation algorithm works in the following way. {First, a random point with value '1' (foreground) in $\bUp^{\rm km}$ is picked up, defined as a central point, and its eight-point neighborhood values are checked. Then, the neighbors that have value '1' to this central point are considered to be in the same region as the central points. The algorithm starts by further searching considering the neighborhood of newfound points. Finally, a shape is defined when there are no other points left with the value '1' around the perimeter, where the outermost points are retrieved as a perimeter. Both perimeter and internal points are defined by $(x,y)$ tuples representing the column and row in which they are allocated in the $\bUp^{\rm km}$ matrix. The perimeter points are assigned to a subset $\mathcal{B}_s \subset \mathcal{B}$ and all the points that define a shape in $\mathcal{A}_s \subset \mathcal{A}$, where $s$ indexes a shape. Then, we associate a value $E_s$ to each region $s$ depicting the total power level within that region, which is} calculated as:
\begin{equation}\label{eq:energyset}
    E_s = \sum\limits_{i \in \mathcal{A}_{s} } \left( \tilde{\bUp}^{\rm NLoS}[x_i,y_i] \right)^2.
\end{equation}
The values $E_s$'s are further stored in the set $\mathcal{E}$ .
The boundaries for this scenario can be seen on the right side of Fig. \ref{fig:segandbwbound}. 

\begin{figure}[t!]
    \centering
    \includegraphics[trim = 0mm 0mm 0mm 0mm, clip, width=.75\linewidth]{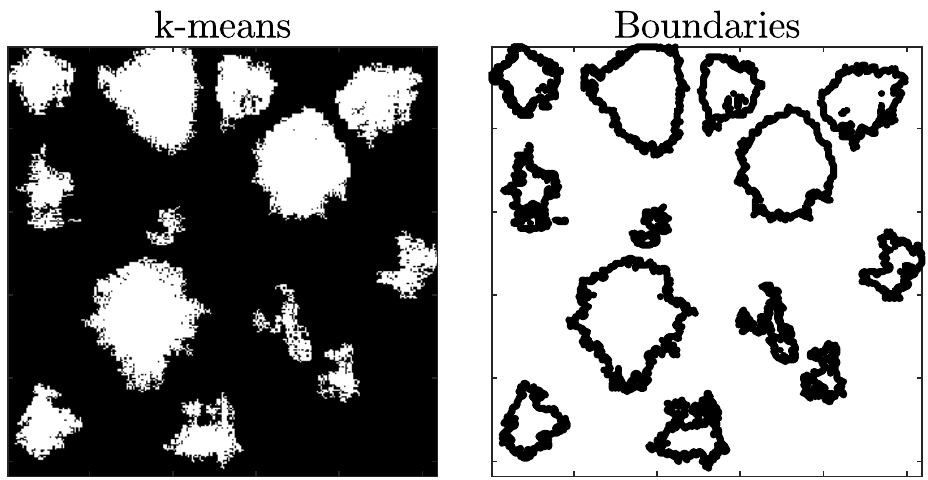}
    \caption{Graphic representation of \textit{left:} the output of the binary k-means clustering, $\bUp^{\rm km}$ and \textit{right:} the estimated boundaries set $\mathcal{B}$ obtained via Moore-Neighbor boundaries estimation.}
    \label{fig:segandbwbound}
\end{figure}

\subsubsection{Sensing and Localization Inference}
This block receives the sets $\mathcal{B}$ and $\mathcal{E}$ and does the inference process. First, the data is clustered with \gls{dbscan} \cite{Ester1996dbscan}. This method randomly chooses a core point by observing the data in $\mathcal{B}$ and maps the neighborhood subject to $\mathtt{minPts}$ and $\epsilon$ constraint parameters, representing the minimum number of points in a cluster and the search radius, respectively. Starting from the core point, the algorithm calculates the distances to all points in the data and assigns as a cluster those that satisfy the parameter constraints. The process is repeated until all the data have been clustered. Points that are sparsely distributed and thus cannot be assigned to a cluster, are considered noise. The indexes of the shapes that compound each cluster are assigned to the $i$-th cluster $\mathcal{I}_i \subset \mathcal{I}$ that maps both $\mathcal{B}$ and $\mathcal{E}$ sets.

We perform a test of the described algorithm on the illustrative scenario in Appendix \ref{appx:illustrative-indoor} with $\mathtt{minPts} =2$ and ${\epsilon =2}$. The algorithm output 63 identified clusters, which is much higher than the true number of \glspl{pe} in the environment. To overcome excessive number of clusters, we exploit the metallic objects' lower reflection loss than the humans, $|\sigma_r| {\ll} |\sigma_{r'}|$, so we can infer that clusters with high power are objects{, while} clusters with {low} powers are noise or background distortion{; consequently, clusters in the middle {represent human positions}. Thus, the $i$-th cluster is considered to be of a human class if it satisfies the following rule over $\mathcal{E}$:}
\begin{equation}\label{eq:classification}
\mathfrak{D}(i) = 
\begin{cases}
1, & {\rm if} \quad {\rm th}_{\min} < \sum_{s \in {\mathcal{I}_i}} \mathcal{E}_s \leq {\rm th}_{\max},\\
0, & {\rm otherwise}.
\end{cases}
\end{equation}
where ${\rm th}_{\min}$ and ${\rm th}_{\max}$ are the minimum and maximum threshold parameters, respectively. If the cluster passes the threshold rule, then the cluster centroid is stored in a new \textit{set of detected humans} $\mathcal{H}$.

To evaluate how {reasonable} the inference is, let $\mathbf{a}\in{\mathbb{R}^+}^{R_{\rm hum}\times 1}$ be the vector containing the differences between the inferred and the ground truth localization of the humans, giving the localization accuracy per human. Its $r$-th entry is calculated as:
\begin{equation}\label{eq:accuracy}
a_{{r}} = \min\left( \left\{ \lVert\mathcal{H}_j - \mathcal{H}^{{\rm gt}}_{{r}} \rVert <d_{\rm th}, 
{j=1,2,\ldots,} 
{|}\mathcal{H}{|} \right\} \right),
\end{equation}
where $\mathcal{H}_j$ denotes the $j$-th element of $\mathcal{H}$ and $\mathcal{H}^{\rm gt}$ is the set of ground-truth humans with {the $r$}-th element denoted as $\mathcal{H}^{\rm gt}_{{r}}$ and whose cardinality is $\lvert\mathcal{H}^{\rm gt}\rvert=R_{\rm hum}$. Furthermore, $\| \cdot \|$ is the Euclidean norm. We adopt as $d_{\rm th} = 1$m the distance threshold for detection via experimentation. The value of accuracy $a_{r}$ can be \textit{null} in case the distance is higher than the one defined by the threshold; in this case, we assume $a_{{r}}=0$. We define the localization accuracy, $\mathrm{LA}$, and the detection rate, $\mathrm{DR}$, as:
\begin{equation}\label{eq:perf-metrics}
    {\rm LA} = \dfrac{1}{R_{\rm hum}}\sum\limits_{i=1}^{R_{\rm hum}} {a}_i, \text{ and } {\rm DR} = \dfrac{1}{R_{\rm hum}}\sum\limits_{i=1}^{R_{\rm hum}} \mathbb{1}({a}_i > 0),
\end{equation}
where $\mathbb{1}$ is the indicator function.

\begin{figure}[t!]
    \centering
\includegraphics[width=.75\linewidth]{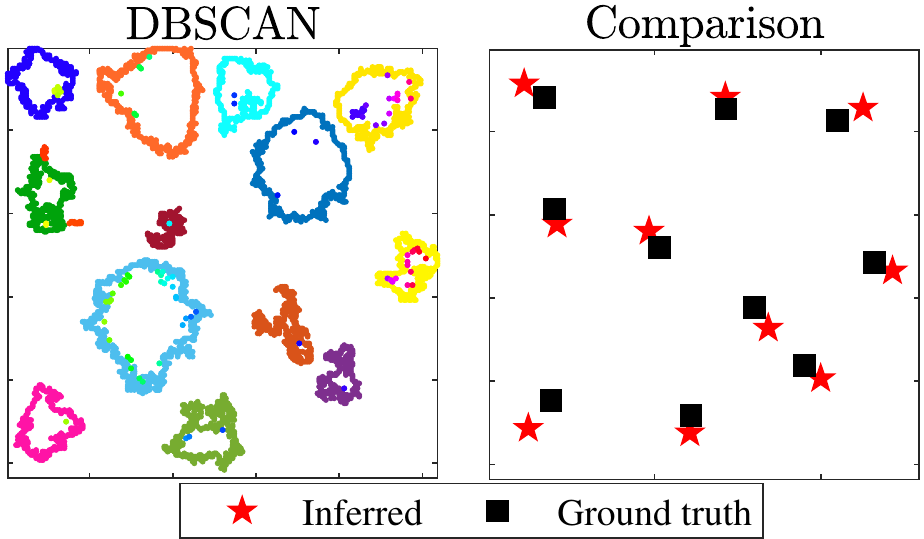}
\caption{{Inference step of OnRmap. \textit{Left}: identified clusters by \gls{dbscan}. {The colors represent each cluster; the large ones are more likely to represent \glspl{pe}, while the clusters with very few elements are likely noise}. \textit{Right}: inferred positions comparison against ground truth.}}
\label{fig:FinalToyExample}
\end{figure}

By empirically setting set ${\rm th}_{\min} = 0.03 \frac{1}{S} \sum \mathcal{E}$ and ${\rm th}_{\max} = 0.8 \max(\mathcal{E})$, Fig. \ref{fig:FinalToyExample} shows how the data were clustered for the illustrative example of Appendix \ref{appx:illustrative-indoor}. Overall, it was identified 67 clusters, where three have very high energy, corresponding to the metallic {objects}, ten with low to medium energy, corresponding to humans, and the rest with {shallow} energy, considered {noise}. To visually assess the quality of the results, we re-plot on the right side of the figure the ground truth points together with the inferred points. The algorithm was capable of inferring the human's positions with localization accuracy from $0.1${m} (best case) to $0.62${m} (worst case) and the average of $0.22${m} in this single realization.

\section{Numerical results}

\begin{figure*}[t]
    \centering
    \includegraphics[trim = 0mm 9mm 0mm 0mm, clip,width = .98\linewidth]{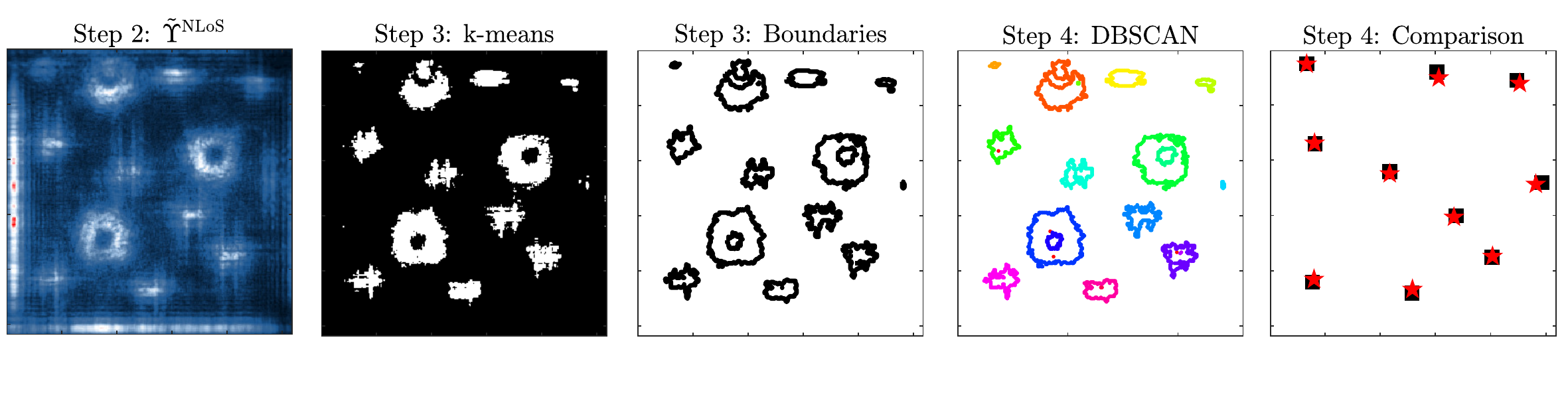}
    \caption{Exemplary application of \gls{our} in the simulation environment from \cite{Vaca-Rubio2021}. In Step 4: \colr{$\star$} is inferred position  and $\smblksquare$ ground-truth.}
    \label{fig:ResultExample}
\end{figure*}


We now evaluate the effectiveness of \gls{our} and compare it with the previous works \cite{Vaca-Rubio2021,vaca2022floor}. For a better evaluation, we consider a more complex indoor scenario as provided by Feko by Altair Engineering ray tracing, as used in \cite{Vaca-Rubio2021}. Different from the system model and toy example presented, this scenario includes more reflections from the \glspl{pe} and reflections from the ground and walls. Despite that, the other parameters are the same as in Appendix \ref{appx:illustrative-indoor} with $R_{\rm obj}=3$ metallic objects and $R_{\rm hum}=10$ humans. \textit{Throughout this section, we focus on showing results for human detection since the detection of \glspl{au} and metallic objects is less challenging.}

\begin{figure}[ht]
    \centering
    \subfloat[{Average localization accuracy and average detection rate, defined in \eqref{eq:perf-metrics}, for different numbers of \glspl{au}.}]{\includegraphics[width = .98\linewidth]{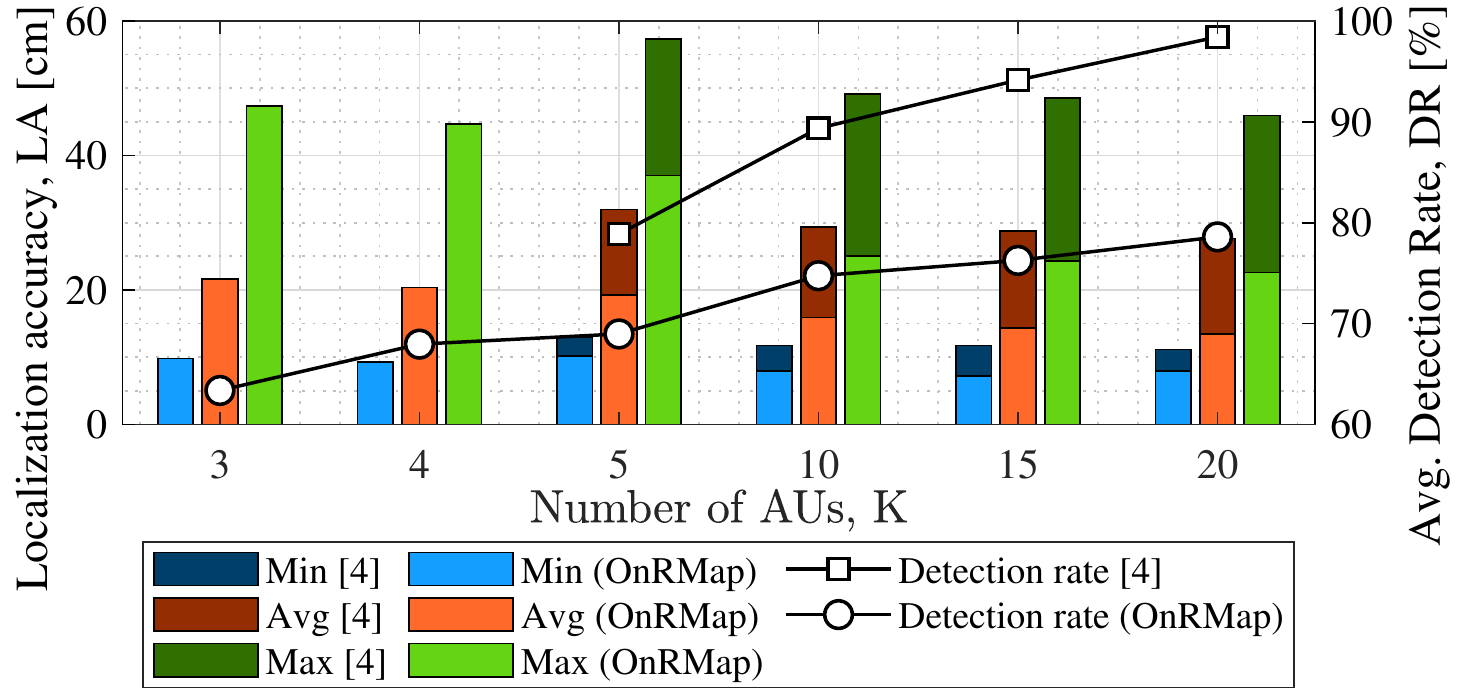}
    \label{fig:performance}}
    \hfil
    \subfloat[CCDF of the number of correctly detected humans for different numbers of \glspl{au}.]{\includegraphics[trim = 0mm 7mm 0mm 0mm, clip, width = .98\linewidth]{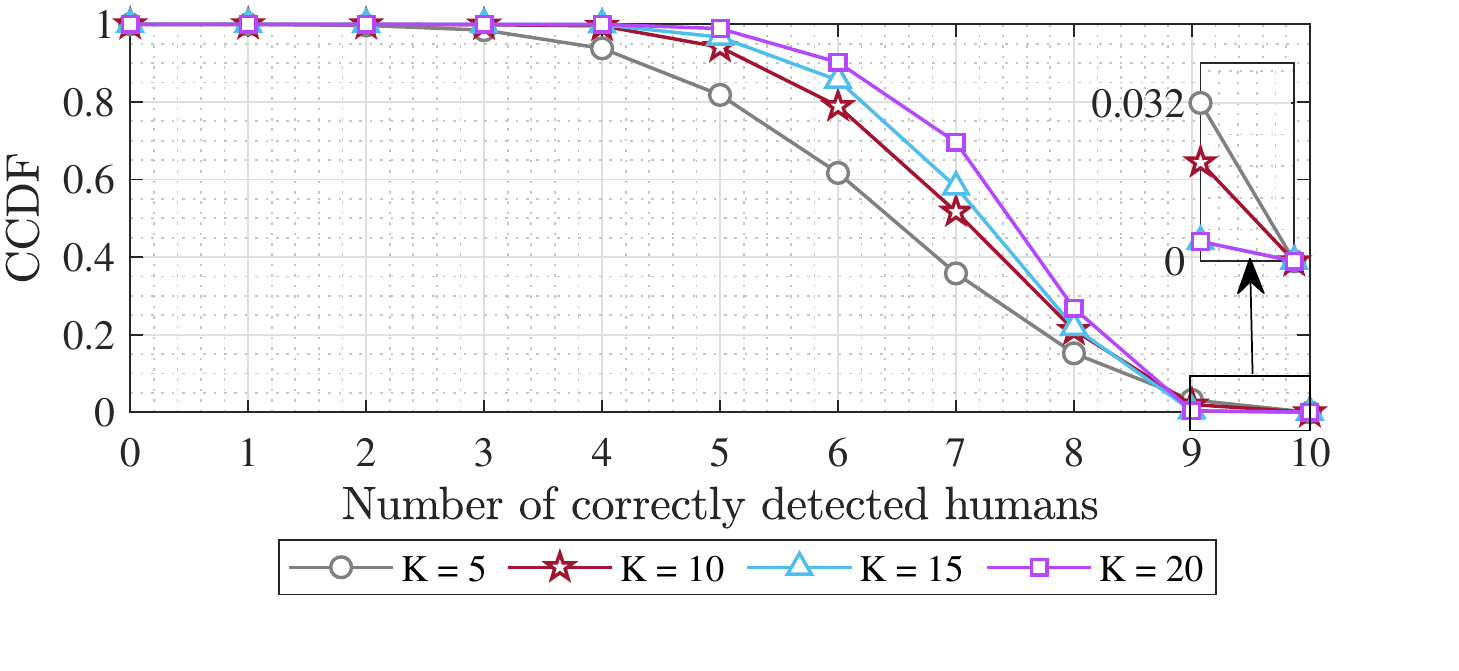}
    \label{fig:probvsmcs}}
	\caption{Performance evaluation for the sensing (detection) and localization of $R_{\rm hum}=10$ humans (\glspl{pe}).}
	\label{fig:all}
\end{figure}

We start by illustrating a step-by-step realization of \gls{our} in Fig. \ref{fig:ResultExample} with $K=20$ \glspl{au}. From the figure, we can note that \gls{our} can perform well even in this more unfavorable scenario, given the increase in multipath components. In Fig. \ref{fig:all}, we better evaluate the performance of \gls{our} by considering 1000 \glspl{mcs}. Fig. \ref{fig:performance} shows the average localization accuracy and the average detection rate of the humans when considering different numbers of \glspl{au}. We compare these results with the ones presented in \cite{Vaca-Rubio2021}. The trade-off in the comparison is that we can achieve higher localization accuracy for all {numbers of \gls{au} configurations ($K$)} at the cost of lower detection rates on average. It is possible to point {out} two causes of the lower detection performance {compared to} \cite{Vaca-Rubio2021}. First, some information of the signals is filtered after \gls{rpca} and does not enter in the inference process, as opposed to \cite{Vaca-Rubio2021} that treats the signals from all \glspl{au}. Second, when the boundaries among \glspl{pe} from different classes are too close, \gls{dbscan} classify them as one cluster and the decision rule considers both as just one type of \gls{pe}. {These are considered the costs of the proposed online method in view of the offline approach. The lack of {\it a priori} information to compensate static \glspl{pe} as in \cite{Vaca-Rubio2021}, compromises the $\mathrm{DR}$ to some extent. However, we argue that the high applicability and the significant improvement in the overall $\mathrm{LA}$ justify the novelty of this work.}

To better understand the impact of the number of \glspl{au} in the detection rate, Fig. \ref{fig:probvsmcs} shows the \gls{ccdf} of the detection in terms of the number of correctly detected humans for different numbers of \glspl{au} $K$. Note that the effect of increasing $K$ is to lower the variance on how many humans $R_{\rm hum}$ we can detect. However, the opposite occurs at the points of nine to ten correctly detected humans, highlighted in Fig. \ref{fig:probvsmcs}. For $K=20$, the probability of a perfect detection (ten humans) is 0.4 \%, and when $K=5$ it achieves 3.2 \%. {A reason for this is that higher $K$ implies more signals reflecting on \glspl{pe} that can be possibly distributed next to the center of the room, while the \glspl{pe} next to the walls are less exposed, and those reflections start to be interpreted as \textit{background} (distortion/noise) by the method.} 

\section{Conclusions}
In this paper, we proposed \gls{our}, an online method for sensing and localization in indoor environments equipped with \gls{lis} systems. The greatest advantages of \gls{our} are due to the fact that it is based on signal processing techniques, making it an online method, that is, it does not rely on offline scanning phases. This makes the method more robust for applications where the environment is constantly changing. However, the online feature comes with the cost of an average lower detection rate. But even so, \gls{our} turns out to have a fairly good location accuracy. Future works may improve the design of \gls{our} to cover the observed weakness and better study its performance.

\appendices
\section{An Illustrative Indoor Scenario}\label{appx:illustrative-indoor}
The parameters of the simulation scenario are summarized in Table \ref{Tab:Simulation}. Note that we consider a scenario with two types of \glspl{pe}: i) {$R_{\rm obj}=3$} cylindrical \textit{metal objects} with polished surfaces $\sigma_r \sim \mathcal{U}[-10,-15]$ {\small 
 dB} and {$R_{\rm hum}=10$} \textit{humans} $\sigma_{r'}\sim \mathcal{U}[-30,-75]$ {\small 
 dB}, totaling $R=13$ \glspl{pe}.

\definecolor{Gray}{gray}{0.95}
\begin{table}[ht]
\centering 
\caption{Simulation parameters}\label{Tab:Simulation}\scriptsize
\begin{tabular}{ll|ll}
\multicolumn{1}{l}{\textbf{Parameter}}                  & \textbf{Value    }               & \multicolumn{1}{l}{\textbf{Parameter}}    & \multicolumn{1}{c}{\textbf{Value }} \\ \hline \rowcolor{Gray}
\multicolumn{2}{c|}{\textbf{\textit{Environment}}}                                     & \multicolumn{2}{c}{{\textbf{\textit{Channel \& System}}}}                                             \\ \hline
\multicolumn{1}{l}{\multirow{2}{*}{\# PEs}} &  $R_{\rm obj}$ = 3 &  \multicolumn{1}{l}{Trans. power} &  $p = 20$ [dBm] \\ 
\multicolumn{1}{l}{}                           & $R_{\rm hum}$ = 10       & \multicolumn{1}{l}{Noise power}  & {$\sigma^2_w = -97$ [dBm] }              \\ \cline{1-2}
\multicolumn{1}{l}{\multirow{3}{*}{Room dim.}} & (y) = 10.36 [m] & \multicolumn{1}{l}{LoS PL} &  $\beta^0_{k,n} = \frac{\lambda}{4\pi d^1_{k,n}}$ \\   
\multicolumn{1}{l}{} & (x) = 10.36 [m] &  \multicolumn{1}{l}{\multirow{2}{*}{NLoS ref. loss}} &  $\sigma_r \sim \mathcal{U}[-10,-15]$ dB \\   
\multicolumn{1}{l}{}                           & (z) = 8 [m]      & \multicolumn{1}{l}{}             & $\sigma_{r'} \sim \mathcal{U}[-30,-75]$ dB \\ \hline
\multicolumn{1}{l}{\multirow{2}{*}{Obj. dim}} &  rad. = 0.43 [m] &  \multicolumn{1}{l}{Carr. freq.} &  $f = 3.75$ [GHz] \\  
\multicolumn{1}{l}{}                           & (z) = 2 [m]      & \multicolumn{1}{l}{Carr. wavel.} & $\lambda = 0.08$ [m]                       \\ \hline
\multicolumn{1}{l}{\multirow{3}{*}{Hum. dim.}} & (y) = 0.3 [m]        & \multicolumn{2}{c}{\cellcolor{Gray} \textbf{\textit{Active users}}} \\ \cline{3-4}
\multicolumn{1}{l}{}                           & (x) = 0.5 [m]       & \multicolumn{1}{l}{\# AUs}        & {$K =9$}          \\ 
\multicolumn{1}{l}{}                           & (z) = 1.7 [m]    & \multicolumn{1}{l}{AUs height}           & $z_{k} = 1.8 $ [m]                         \\ \cline{1-2}
\multicolumn{1}{l}{LIS pos.}                   & $z=8$  [m]          &              &                                            \\
\multicolumn{1}{l}{LIS elem.}                  & $N_x=N_y = 259$  &              &                                            
\end{tabular}
\end{table}

\bibliographystyle{IEEEtran}
\bibliography{ref.bib}

\end{document}